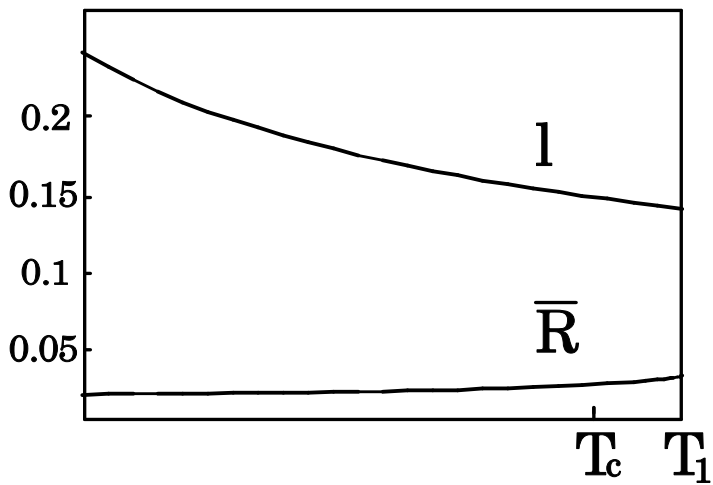

Fig. 1

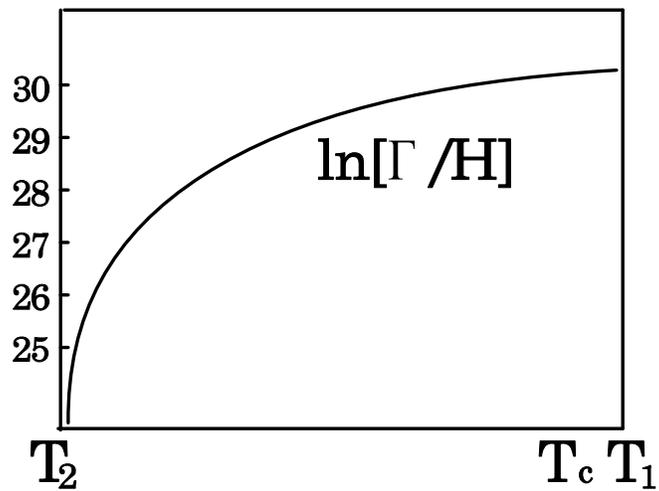

Fig. 3

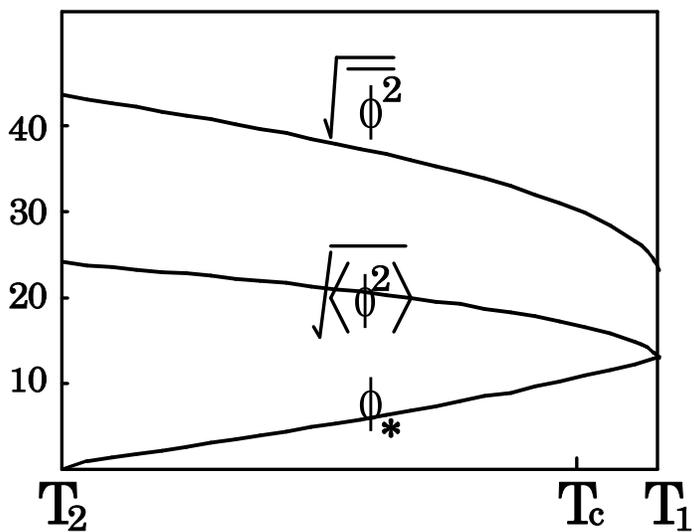

Fig. 2

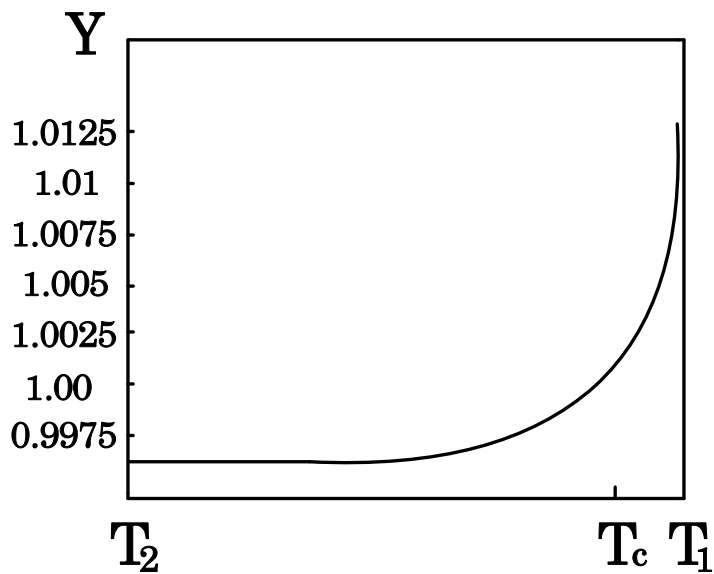

Fig. 4

# Thermal Fluctuations in Electroweak Phase Transition

Tetsuya Shiromizu
*Department of Physics, Kyoto University, Kyoto 606-01, Japan*

Masahiro Morikawa
*Department of Physics, Ochanomizu University, Tokyo 112, Japan*

Jun'ichi Yokoyama
*Uji Research Center, Yukawa Institute for Theoretical Physics*
*Kyoto University, Uji, Kyoto 611, Japan*

### Abstract

We estimate the amplitude of thermal fluctuations by calculating the typical size of subcritical bubbles in cosmological electroweak phase transition and show that this thermal fluctuation effect drastically changes dynamics of the phase transition from the ordinary first order type with supercooling. From this fact, we conclude that the standard electroweak baryogenesis scenario associated with such a first order transition does not work in the minimal standard model in certain conditions.

Recently electroweak baryogenesis scenario has been actively investigated [1] [2]. In this scenario the phase transition must be of first order with supercooling in order to attain a non-equilibrium state for baryon number non-conserving processes. In one-loop appoxmation, the finite temperature effective potential always yields a barrier between the false and the true vacua, *i.e.*, a first order phase transition [3]. However, if thermal fluctuation is too large, the perturbation scheme breaks down and its prediction of a first-order phase transition becomes suspect. Although the dynamics of the phase transition is of primary importance in such a cosmological context, the analyses so far done have mainly been confined in other aspects, namely, spontaneous baryon-number generation in thick walls [1] and charge transport mechanism in thin walls [2]. In fact, if the transition is not of first order with supercooling, then these mechanisms do not work, invalidating the entire scenario. In the present paper we consider effects of thermal fluctuations on the electroweak phase transition paying attention to the role of subcritical bubbles.

So far Gleiser et al. [4] and Dine et al. [5] have estimated the amplitude of thermal fluctuations and obtained the result that they are dominant if $m_H > 80$GeV [4] and therefore the critical bubble cannot be borne due to inhomogeneities of background fields. As the experimental constraint is $m_H > 60$GeV [6], the parameter range allowing electroweak baryogenesis is still open. However, in their analyses, they assumed that the spatial scale of the thermal fluctuation is equal to the correlation length, that is, the inverse mass scale. In the microscopic level, the true nature of the dominant thermal fluctuation would be a perpetual creation and annihilation of spherical subcritical bubbles. Thus one should identify the typical size of the bubbles with that estimated by the statistical ensemble averaging. What we would like to claim here is that the amplitude of the thermal fluctuation sensitively depends on the size of the subcritical bubble. A small change of the size results in a drastic change of the phase transition.

To derive the characteristic scale of a subcritical bubble we shall construct effective Lagrangian for the size of the bubble $R(t)$. In high temperature phase, the motion for the wall of asymmetric subcritical bubble is bounded from the above and then one can obtain the finite thermal average scale $\overline{R}$ of $R(t)$. On the other hand, in the low temperature phase, symmetric subcritical bubble is bounded from the above also and the size is almost the same as the above $\overline{R}$. The correlation length is the maximum scale of thermal correlation and dose not necessarily coincide with the size of the subcritical bubble in general. This means that $\overline{R}$ is smaller than the correlation length and that the amplitude of thermal fluctuation will increase, invalidating perturbation analysis. Intuitively, this means that one cannot find the distinction between symmetric vacuum and asymmetric vacuum and that the spontaneous critical bubble nucleation out of the homogeneous symmetric state, which is usually assumed, cannot happen.

In this paper we concentrate on the minimal standard model. The effective potential with one-loop and finite temperature correction [3] is given by

$$V_T(\phi) = D(T^2 - T_2^2)\phi^2 - ET\phi^3 + \frac{1}{4}\lambda_T\phi^4, \tag{1}$$

where

$$D = \frac{1}{24}\left[6(\frac{m_W}{\sigma})^2 + 3(\frac{m_Z}{\sigma})^2 + 6(\frac{m_t}{\sigma})^2\right] \sim 0.169, \qquad (2)$$

$$E = \frac{1}{12\pi}\left[6(\frac{m_W}{\sigma})^3 + 3(\frac{m_Z}{\sigma})^3\right] \sim 0.00965, \qquad (3)$$

and

$$\lambda_T = \lambda - \frac{1}{16\pi^2}\left[\sum_B g_B(\frac{m_B}{\sigma})^4 \ln\left(\frac{m_B^2}{c_B T^2}\right) - \sum_F g_F(\frac{m_F}{\sigma})^4 \ln\left(\frac{m_F^2}{c_F T^2}\right)\right] \sim 0.0350, \qquad (4)$$

where we have taken $m_W = 80.6\text{GeV}$, $m_Z = 91.2\text{GeV}$, $m_t = 174\text{GeV}$ and $\sigma = 246\text{GeV}$ [7]. In this case $T_2 = \sqrt{(m_H^2 - 8B\sigma^2)/4D} \sim 92.69\text{GeV}$ and the critical temperature at which the hight of the potential at two local minima, $\phi_- := 0$ and $\phi_+ := \frac{3ET}{2\lambda_T}[1 + \sqrt{1 - \frac{8\lambda_T D}{9E^2 T^2}(T^2 - T_2^2)}]$, degenerates is $T_c = T_2/\sqrt{1 - E^2/\lambda_T D} \sim 93.43\text{GeV}$, where

$$B = \frac{1}{64\pi^2}\left[6(\frac{m_W}{\sigma})^4 + 3(\frac{m_Z}{\sigma})^4 - 12(\frac{m_t}{\sigma})^4\right] \sim -0.00456. \qquad (5)$$

The temperature at which the global minimum appears is $T_1 = T_2/\sqrt{1 - 9E^2/8\lambda_T D} \sim 93.52\text{GeV}$.

We have also chosen the minimal value $m_H = 60\text{GeV}$ for Higgs mass permitted from experimental constraint. As the thermal fluctuation is an increasing function of the Higgs mass, it is sufficient for our purpose if thermal fluctuation is shown to be too large with $m_H = 60\text{GeV}$. On the other hand, in order to avoid baryon number destruction due to sphaleron-mediated processes after the phase transition, we need $m_H < 55\text{GeV}$ in one-loop perturbation theory [8]. Further, Shaposhnikov has obtained the result $m_H < 100\text{GeV}$ in lattice calculation [9]. Recently, the loop correction for the sphaleron transition rate have been estimated and in this case $m_H < 63\text{GeV}$ [10]. Thus, the possibility for the explanation of the present net baryon-number remains at present time.

As we have already mentioned, the typical size of thermal fluctuation is a calculable quantity. Since one cannot use the instanton method for them, we must invent new method. The subcritical bubble which minimizes the free energy would have the $O(3)$ symmetry. Hence in the Lagrangian,

$$L = \int d^3x \left[\frac{1}{2}\dot{\phi}^2 - \frac{1}{2}(\nabla\phi)^2 - V_T(\phi)\right], \qquad (6)$$

we adopt the following ansatz for the bubble configuration:

$$\phi = \phi_+ \exp\left[-\frac{r^2}{R(t)^2}\right], \qquad r := |\mathbf{x}|, \qquad (7)$$

where $R(t)$ is the size of subcritical bubble and is time dependent. Without any definite argument, previous authors have chosen the correlation length $\ell(T) = 1/\sqrt{2D(T^2 - T_2^2)}$ as the typical size $R(t)$. However, in the microscopic level, the bubble has dynamics. Inserting (7) into the above Lagrangian (6) yields an effective Lagrangian for $R(t)$:

$$L_{eff} = \frac{15\pi^{3/2}}{16\sqrt{2}}\phi_+^2 R\dot{R}^2 - \frac{3\pi^{3/2}}{4\sqrt{2}}\phi_+^2 R - \frac{\pi^{3/2}}{2\sqrt{2}}D(T^2 - T_2^2)\phi_+^2 R^3 + \frac{\pi^{3/2}}{3\sqrt{3}}ET\phi_+^3 R^3 - \frac{\pi^{3/2}}{32}\lambda_T \phi_+^4 R^3. \qquad (8)$$

Applying the variational principle on $R(t)$ we obtain the equation of motion for $R(t)$:

$$\frac{d^2 R}{dt^2} + \frac{1}{2R}\left(\frac{dR}{dt}\right)^2 + \frac{2}{5}\frac{1}{R} + \frac{4}{5}D(T^2 - T_2^2)R - \frac{8\sqrt{2}}{15\sqrt{3}}ET\phi_+ R + \frac{1}{10\sqrt{2}}\phi_+^2 \lambda_T R = 0. \qquad (9)$$

In the vicinity of the turning point $|dR/dt| \ll 1$, one can omit the second term of the left-hand-side and then we get an energy equation by multiplying $dR/dt$ and integrating over $t$, which reads

$$\frac{1}{2}\left(\frac{dR}{dt}\right)^2 + \mathcal{U}(R, \phi_+, T) = E_0, \qquad (10)$$

with

$$\mathcal{U}(R,\phi_+,T) := \frac{2}{5}\ln\frac{R}{R_c} + \left[\frac{2}{5}D(T^2-T_2^2) - \frac{4\sqrt{2}}{15\sqrt{3}}ET\phi_+ + \frac{1}{20\sqrt{2}}\lambda_T\phi_+^2\right]R^2 \tag{11}$$

where $E_0$ is a constant of motion and $R_c := \sqrt{E_0/U(\phi_+,T)}$ is the turning radius. The potential $\mathcal{U}(R,\phi,T)$ is a monotonically increasing function of $R$ and therefore the wall motion is restricted from the above in the high temperature phase($T > T_c$). The existence of the upper limit will give the typical size of the subcritical bubble.

What we would like to obtain is the typical size of a subcritical bubble. In order to estimate it we must construct the Hamiltonian which describes the dynamics for $R(t)$. The canonical momentum is defined by

$$P := \frac{\partial L_{eff}}{\partial \dot{R}} = \frac{15\pi^{3/2}}{8\sqrt{2}}\phi_+^2 \dot{R}R. \tag{12}$$

The effective Hamiltonian then becomes

$$\begin{aligned}H_{eff}(R,P) &:= P\dot{R} - L_{eff} \\
&= \frac{4\sqrt{2}}{15\pi^{3/2}\phi_+^2}\frac{P^2}{R} + \frac{3\pi^{3/2}}{4\sqrt{2}}\phi_+^2 R + \left[\frac{\pi^{3/2}}{2\sqrt{2}}D(T^2-T_2^2)\phi_+^2 - \frac{\pi^{3/2}}{3\sqrt{3}}ET\phi_+^3 + \frac{\pi^{3/2}}{32}\lambda_T\phi_+^4\right]R^3 \\
&=: \frac{4\sqrt{2}}{15\pi^{3/2}\phi_+^2}\frac{P^2}{R} + W(R) \\
&=: \frac{4\sqrt{2}}{15\pi^{3/2}\phi_+^2}\frac{P^2}{R} + \frac{R}{\alpha} + \frac{R^3}{\beta}.\end{aligned} \tag{13}$$

Thus the thermal average of $R$ is given by

$$\begin{aligned}\overline{R} &:= \frac{\int \frac{dPdR}{2\pi} R \exp\left[-\frac{H_{eff}}{T}\right]}{\int \frac{dPdR}{2\pi} \exp\left[-\frac{H_{eff}}{T}\right]} \\
&= \frac{\int_0^\infty dR\, R^{3/2}\exp\left[-\frac{W(R)}{T}\right]}{\int_0^\infty dR\, R^{1/2}\exp\left[-\frac{W(R)}{T}\right]} \\
&\sim \frac{\int_0^\infty dR\, R^{3/2}\exp\left[-\frac{R}{\alpha}\right]}{\int_0^\infty dR\, R^{1/2}\exp\left[-\frac{R}{\alpha}\right]} \\
&\sim \frac{3}{2}\alpha.\end{aligned} \tag{14}$$

As $\alpha \sim 0.015 \ll 1$ and $\alpha^3/\beta \sim 0.5$, the last appoxmation is certainly justified. Further, one must treat carefully the measure. As

$$d\phi = \frac{2\phi_+ r^2}{R^3}e^{-\frac{r^2}{R^2}}dR \tag{15}$$

and

$$dp_\phi = -\frac{32\sqrt{2}r^2}{15\pi^{3/2}R^5\phi_+}e^{-\frac{r^2}{R^2}}PdR + \frac{16\sqrt{2}r^2}{15\pi^{3/2}R^4\phi_+}e^{-\frac{r^2}{R^2}}dP, \tag{16}$$

the original measure becomes

$$\int d^3x\, d\phi\, dp_\phi = dRdP \int d^3x \frac{\partial(\phi,p_\phi)}{\partial(R,P)} = dRdP. \tag{17}$$

Thus, naively used measure in eq.(14) is justified. Moreover, since the three-dimensional volume integral is done in eq. (17), weight associated with three volume has already been taken into account. The temperature dependence of eq.(14) is depicted in Fig.1. As one can easily see, the above result gives an *upper* bound for $\overline{R}$. In fact, at $T = T_c$ and $T = T_1$, we numerically obtain the more accurate result that $\overline{R}(T=T_c) = 0.012\text{GeV}^{-1}$ and $\overline{R}(T=T_1) = 0.0202\text{GeV}^{-1}$, respectively. In Fig. 1, we can easily find $\overline{R}(T) < \ell(T)$ as expected.

Next, let us estimate the amplitude of the thermal fluctuation on the scale $\overline{R}$, adopting the trial function

$$\phi = \phi_A \exp\left[-\frac{r^2}{\overline{R}^2}\right]. \tag{18}$$

Then the free energy becomes

$$F_0(\phi_A, T) = \left[\frac{3\pi^{3/2}}{4\sqrt{2}}\overline{R} + \frac{\pi^{3/2}}{2\sqrt{2}}D(T^2 - T_2^2)\overline{R}^3\right]\phi_A^2 - \frac{\pi^{3/2}}{3\sqrt{3}}ET\phi_A^3\overline{R}^3 + \frac{\pi^{3/2}}{32}\lambda_T\phi_A^4\overline{R}^3. \tag{19}$$

Thus the RMS amplitude of $\phi$ at the symmetric vacuum is

$$\sqrt{\overline{\phi^2}} = \sqrt{\frac{\int d\phi_A \phi_A^2 e^{-\frac{F_0(\phi_A,T)}{T}}}{\int d\phi_A e^{-\frac{F_0(\phi_A,T)}{T}}}} \sim \frac{\phi_+}{\sqrt{3 + \frac{16D(T^2-T_2^2)T^2}{\pi^3 \phi_+^4}}}. \tag{20}$$

Its temperature dependence is depicted in Fig.2. In the same way as in the case of $\overline{R}$, the above approximation result gives the *lower* bound for $\sqrt{\overline{\phi^2}}$. At $T = T_c$ and $T = T_1$, one obtains numerically $\sqrt{\overline{\phi^2}}(T = T_c) = 36.2\text{GeV}$ and $\sqrt{\overline{\phi^2}}(T = T_1) = 28.2\text{GeV}$, respectively. This exceeds the first reflection point $\phi_* = \frac{ET}{\lambda_T} - \sqrt{\frac{E^2T^2}{\lambda_T^2} - \frac{2D(T^2-T_2^2)}{3\lambda_T}}$, which implies that the perturbation theory breaks down and that one can no longer conclude that electroweak phase transition is of first order. For the comparison we calculate the RMS by the same treatment as Dine et al [5]. According to their analysis, it becomes

$$\langle\phi^2\rangle_{k<1/\overline{R}} \sim \frac{T}{2\pi^2 \overline{R}} \sim \frac{\phi_+^2}{4\sqrt{2}\pi^{1/2}} \tag{21}$$

and then $\sqrt{\langle\phi^2\rangle} \sim \phi_+/3$. This value is slightly smaller than eq. (20), but exceeds $\phi_*$ in the most range of temperature as one can see in Fig. 2.

To obtain the general view of the phase transition we must examine the dynamics of the symmetric subcritical bubbles. In the low temperature, symmetric bubbles are also always bounded, and this gives the typical scale of the subcritical bubble. We adopt the trial function for a symmetric bubble in the asymmetric phase as follows:

$$\phi = \phi_+\left[1 - \exp\left[-\frac{r^2}{R(t)^2}\right]\right]. \tag{22}$$

The same procedure as the above discussion yields the energy equation near the turning point $R_c$:

$$\frac{1}{2}\left(\frac{dR}{dt}\right)^2 + \tilde{\mathcal{U}}(R, \phi_+, T) = E_0, \tag{23}$$

where

$$\tilde{\mathcal{U}}(R, \phi_+, T) := \frac{2}{5}\ln\frac{R}{R_c} - \left[\frac{4\sqrt{2}}{5}(2 - \frac{1}{2\sqrt{2}})D(T^2 - T_2^2) - \frac{4\sqrt{2}}{5}(3 - \frac{3}{2\sqrt{2}} - \frac{1}{3\sqrt{3}})ET\phi_+ \right.$$
$$\left. + \frac{2\sqrt{2}}{5}(4 - \frac{1}{8} - \frac{3}{\sqrt{2}} + \frac{4}{3\sqrt{3}})\lambda_T\phi_+^2\right]R^2, \tag{24}$$

and $E_0$ is the constant of motion. Note that the above energy equation holds in the case in which turning point exists. In the low temperature phase($T < T_c$), the potential $\tilde{\mathcal{U}}(R, \phi_+, T)$ is a monotonically increasing function of $R$. To estimate the typical size we construct the effective Hamiltonian for the variable $R(t)$ and its canonically conjugate momentum $P(t)$:

$$H_{eff}(R, P) = \frac{4\sqrt{2}}{15\pi^{3/2}\phi_+^2}\frac{P^2}{R} + \frac{3\pi^{3/2}}{4\sqrt{2}}\phi_+^2 R - \pi^{3/2}(2 - \frac{1}{2\sqrt{2}})D(T^2 - T_2^2)\phi_+^2 R^3 + \pi^{3/2}(3 - \frac{3}{2\sqrt{2}} + \frac{1}{3\sqrt{3}})ET\phi_+^3 R^3$$
$$- \pi^{3/2}(1 - \frac{1}{32} - \frac{3}{4\sqrt{2}} + \frac{1}{3\sqrt{3}})\lambda_T\phi_+^4 R^3. \tag{25}$$

We obtain the same result

$$\overline{R} \sim \frac{2\sqrt{2}T}{\pi^{3/2}\phi_+^2} \tag{26}$$

as the asymmetric bubble in the same approximation level. By the same argument with the asymmetric bubbles, this gives an upper bound for numerical results.

Let us clarify the total dynamics of electroweak phase transition. The following argument is based on the work in ref. [11]. Although it is a rough estimate, the results give us qualitative picture which is necessary for understanding the phase transition. Then transition rate becomes

$$\Gamma_{[0\to+]} = m(T)\exp\left[-\frac{F_0(\phi_+)}{T}\right], \tag{27}$$

and

$$\Gamma_{[+\to 0]} = m(T)\exp\left[-\frac{F_+(\phi_+)}{T}\right], \tag{28}$$

with $m(T) = \sqrt{2D(T^2 - T_2^2)}$, where

$$F_0(\phi_+) = \frac{3}{2}T + \frac{1}{\pi^3}\left[8D(T^2 - T_2^2)\phi_+^{-4} - \frac{16\sqrt{2}}{3\sqrt{3}}ET\phi_+^{-3} + \frac{1}{\sqrt{2}}\lambda_T\phi_+^{-2}\right]T^3, \tag{29}$$

and

$$F_+(\phi_+) = \frac{3}{2}T - \frac{16\sqrt{2}}{\pi^3}\left[(2 - \frac{1}{2\sqrt{2}})D(T^2 - T_2^2)\phi_+^{-4} + \pi^{3/2}(3 - \frac{3}{2\sqrt{2}} - \frac{1}{3\sqrt{3}})ET\phi_+^{-3}\right.$$
$$\left. - \pi^{3/2}(1 - \frac{1}{32} - \frac{3}{4\sqrt{2}} + \frac{1}{3\sqrt{3}})\lambda_T\phi_+^{-2}\right]T^3. \tag{30}$$

For the more detailed analysis of phase transition the above formulae are not sufficient. In particular, note that the prefactor $m(T)$ is determined from the dimensional argument although this term is important when we determine the freeze out temperature $T_{fo}$. The master equations for the number density of the symmetric (asymmetric) state, $N_0$ ($N_+$), are

$$\frac{dN_0}{dt} = -\Gamma_0 N_0 + \Gamma_+ N_+, \tag{31}$$

and

$$\frac{dN_+}{dt} = -\Gamma_+ N_+ + \Gamma_0 N_0, \tag{32}$$

respectively. The behaviour of the both transition rates can be seen in Fig.3, where we can find that $T_{fo}$ is barely larger than $T_2$. However, this results is mainly due to the prefactor $m(T)$. As the both transition rates are much larger than $H = \dot{a}/a$ during the phase transition, $T_{fo}$ is probably lower than $T_2$ if one can calculate the exact transition rate. For example, if we substitute $m(T)$ by $T$, $T_{fo}$ becomes lower than $T_2$. Thus the fraction of false vacuum for true vacuum $Y = N_0/N_+$ has the equilibrium value at $T = T_2$:

$$Y = Y^{eq} = \frac{N_0^{eq}}{N_+^{eq}} = \frac{\Gamma_+}{\Gamma_0} = \exp\left[-\frac{F_+ - F_0}{T}\right], \tag{33}$$

Its temperature dependence is seen in Fig.4. When the temperature becomes $T_1$, the asymmetric vacuum appears. The fraction which the asymmetric vacuum occupies will be almost zero at this moment. The transition rate from asymmetric to symmetric vacuum exceeds during a small temperature difference from $T = T_1$. When the temperature becomes the critical temperature, the fraction $Y$ becomes unity. If the freeze-out temperature is higher than $T_2$, after the temperature becomes the freeze-out temperature the symmetric region will collapse by surface tension. The phase transition cannot be accomplished by critical bubbles. If the freeze out temperature is lower than $T_2$, the phase transition becomes of the second order type. Here we remark that the last attempt exists and some new aspect have been obtained in the case with light Higgs mass($m_H \leq 53$GeV) by considering the kinematics of the subcritical bubble [12]. For such a case, destruction of symmetric bubble by thermal noise becomes dominant.

In this paper, we investigated the strength of the thermal fluctuation and the structure of the phase transition. Our main focus is the estimation of the size of sub-critical bubbles. According to our calculation, the typical size is smaller than the correlation length and the amplitude of thermal fluctuation increases compared with the previous estimation. For experimentally allowed values of Higgs mass, the amplitude of thermal fluctuation always exceeds the first reflection point and then perturbation scheme is not guaranteed. At the critical temperature, the fraction which symmetric and asymmetric vacua occupy is almost the same and therefore the critical bubble cannot be nucleated. Note that Borrill and Gleiser have obtained the results that the two phase mixing is completely attained in numerical experiment [13] and our discussion might be supported by this.

When we obtain effective Lagrangian for the size $R(t)$ we used the thick wall ansatz. In this case, one can easily confirm that the typical size $\overline{R}$ of subcritical bubble is comparable with the thickness of the wall. Thus, the thin wall is not suitable for subcritical bubbles.

In the microscopic level, the subcritical bubble expands and then collapses and any departure from thermal equilibrium is not provided. Thus, the baryon number cannot be generated in minimal standard model. We conclude that the baryogenesis is quite difficult in the minimal standard model, even apart from the fact that CP-violation source is in Kobayashi-Maskawa phase only which is quite small. Thus, we need to investigate "beyond the standard model" and this work will be reported in the near future.

### Acknowledgement

We would like to thank H. Sato and M. Sasaki for their useful comments. TS thanks T. Tanaka for discussions. MM thanks Kurata foundation for financial support. This work was partly supported by Grant-in-Aid for Scientific Research Fellowship, No. 2925 (TS) and by the Scientific Research Fund of Ministry of Education, Science, and Culture, No. 06740216 (JY).

## Figure Captions

- Fig.1: The thermal correlation length $\ell$ and the typical size $\overline{R}$ of subcritical bubbles. The unit of the vertical axis is GeV$^{-1}$.

- Fig.2: The RMS of thermal fluctuations. The unit of the the vertical axis is GeV.

- Fig.3: Transition rates. Both rates degenerate in this figure.

- Fig.4: The fraction of the symmetric phase.